\documentclass{Interspeech2024}

\usepackage{amsmath,graphicx}
\usepackage{multirow,booktabs}

\usepackage{comment}
\usepackage[draft]{todonotes}   % notes showed
\usepackage{soul}
\makeatletter
\newcommand{\thickhline}{%
    \noalign {\ifnum 0=`}\fi \hrule height 1pt
    \futurelet \reserved@a \@xhline
}
\makeatother

\interspeechcameraready

\title{Prompting Whisper for QA-driven Zero-shot End-to-end Spoken \\Language Understanding}

\name{Mohan Li, Simon Keizer, Rama Doddipatla}
\address{Cambridge Research Laboratory, Toshiba Europe Ltd, Cambridge, UK}
\email{\{mohan.li, simon.keizer, rama.doddipatla\}@toshiba.eu}

\begin{document}

\maketitle
 
\begin{abstract}
Zero-shot spoken language understanding (SLU) enables systems to comprehend user utterances in new domains without prior exposure to training data. Recent studies often rely on large language models (LLMs), leading to excessive footprints and complexity. This paper proposes the use of Whisper, a standalone speech processing model, for zero-shot end-to-end (E2E) SLU. To handle unseen semantic labels, SLU tasks are integrated into a question-answering (QA) framework, which prompts the Whisper decoder for semantics deduction. The system is efficiently trained with prefix-tuning, optimising a minimal set of parameters rather than the entire Whisper model. We show that the proposed system achieves a 40.7\% absolute gain for slot filling (SLU-F1) on SLURP compared to a recently introduced zero-shot benchmark. Furthermore, it performs comparably to a Whisper-GPT-2 modular system under both in-corpus and cross-corpus evaluation settings, but with a relative 34.8\% reduction in model parameters.

\end{abstract}
%\noindent\textbf{Index Terms}: zero-shot learning, spoken language understanding, Whisper, prefix-tuning

%\vspace{2mm}

\section{Introduction}
\label{section:intro}

Spoken Language Understanding (SLU) is a fundamental technology in conversational AI, dedicated to converting spoken utterances into semantic elements such as user intents, entities and emotions \cite{tur2011spoken}. The training of SLU models involves expensive data collection, requiring crowdsourced speech recording and expert annotation. This has led to increased research targeting SLU within zero-shot learning scenarios, which aims to develop a generalised system capable of predicting unseen semantic labels without additional training on annotated data.

Previous studies primarily focused on text-based zero-shot natural language understanding (NLU) \cite{xia2018zero, bapna2017towards, shah2019robust, lee2019zero}, which processes transcripts produced by an automatic speech recognition (ASR) model to create a modular solution to zero-shot SLU. Among these studies, the prompt-based question-answering (QA) framework \cite{du2021qa, mehri2021gensf, li2023generative} has gained popularity, driven by the recent advancements in generative large language models (LLMs) \cite{devlin2019bert, radford2019language, raffel2020exploring}. This approach involves crafting a descriptive question for each semantic label (e.g. ``\emph{What is the name of the game?}'' for the slot type {\fontfamily{qcr}\selectfont game\_name}) and prompting an LLM to produce answers (slot values) that fulfill the SLU task. Unlike traditional classification or sequence-tagging methods \cite{mesnil2014using, louvan2020recent} that require predefined output classes, the QA framework offers flexibility in accommodating unseen semantics, relying solely on generating relevant questions for the new labels.

Over the years, SLU has evolved from modular systems to more efficient end-to-end (E2E) models \cite{lugosch2019speech, serdyuk2018towards, haghani2018audio}. However, research on zero-shot E2E SLU remains limited. A recent study introduces a zero-shot audio-to-intent classification framework \cite{elluru2023generalized}, identifying unseen intents by comparing the speech embedding similarity between the test utterance and enrollment samples. Furthermore, \cite{sun2023knowledge} presents a knowledge-aware audio-grounded (KA2G) generative framework for zero-shot slot filling. Similar to modular systems, KA2G incorporates ASR and NLU components to implement the QA-style slot value generation. In this model, these components are aligned in terms of decoder states to facilitate E2E training and inference.

In tandem with the rise of LLMs, advanced speech-to-text (STT) models are emerging, represented by OpenAI's Whisper \cite{radford2023robust}. Developed using a substantial volume of audio-transcript pairs from the internet, Whisper demonstrates outstanding performance in ASR, speech translation (ST), and has been extended to other SLU tasks \cite{wang2023whislu, meeus2023whisper, porjazovski2023advancing} under supervised learning settings. Particularly, these implementations do not depend on external language models. In this paper, we aim to unveil Whisper's capabilities in simultaneously addressing two zero-shot SLU challenges: intent classification and slot filling. Our contributions are summarised as follows:
\begin{itemize}
    \item We investigate the application of the Whisper model for zero-shot E2E SLU, and show that it significantly outperforms the existing baseline on the SLURP benchmark \cite{bastianelli2020slurp}.
    \item To enable zero-shot functionalities, we reformulate the SLU tasks as QA problems and accordingly adapt the Whisper model through decoder-prompting and prefix-tuning.
    \item The proposed system achieves competitive performance compared to a Whisper-GPT-2 pipeline structure, while maintaining a substantially smaller footprint.
    \item We conduct cross-corpus zero-shot evaluations to validate the generalisation of the proposed system, trained on SLURP and tested on FSC \cite{lugosch2019speech} and SmartLight \cite{saade2019spoken} datasets.
\end{itemize}

\begin{figure*}[t]
\centering
  \centerline{\includegraphics[width=\textwidth]{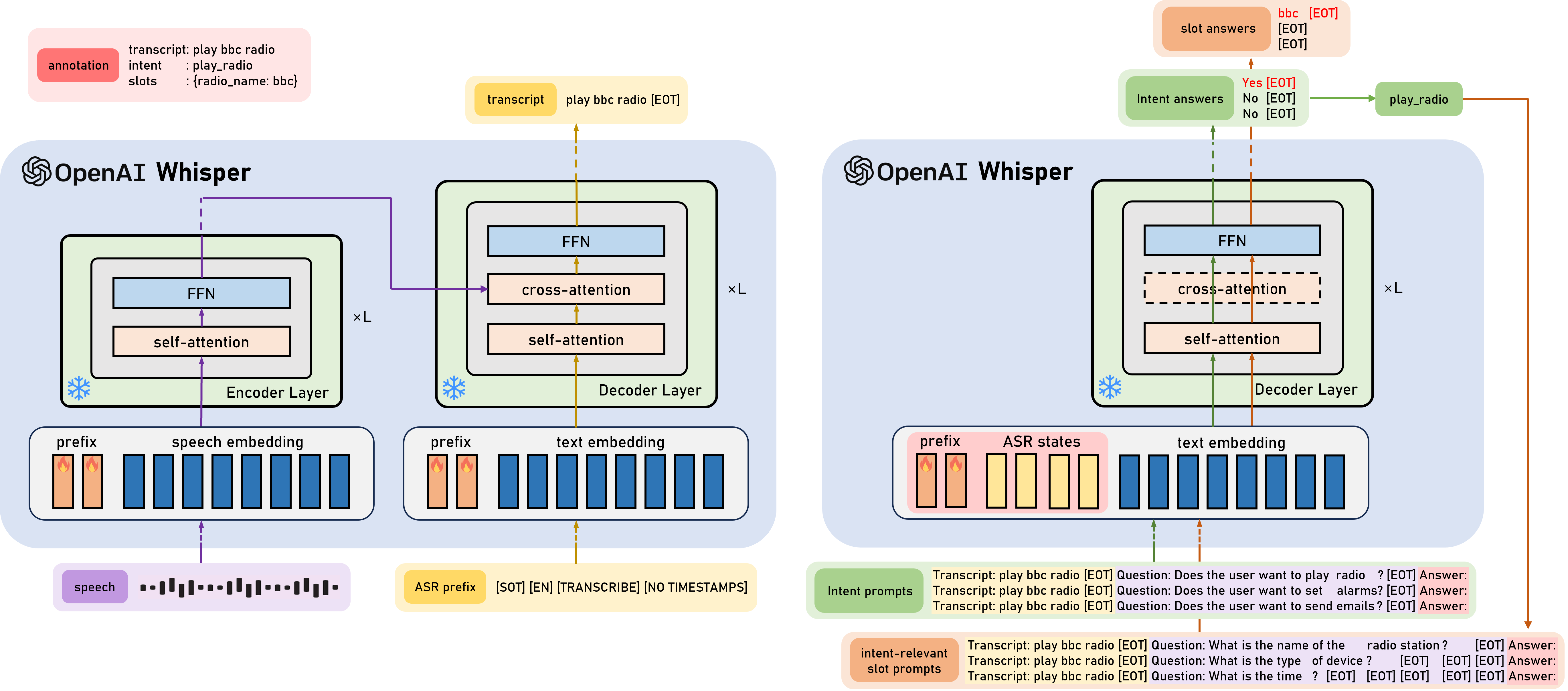}}
\caption{The workflow of the proposed ZS-Whisper-SLU system. The SLU tasks are performed in three stages: ASR (left), intent classification and intent-restricted slot filling (right). The Whisper model is frozen and adapted with prefix-tuning.} 
\label{fig:zs-whisper-slu}
\vspace{-2mm}
\end{figure*}

\section{Proposed Zero-shot Whisper SLU}
\label{section:zs-whisper-slu}

In this section, we provide an overview of the Whisper model and introduce its integration into the prompt-based QA framework. We further elaborate the use of prefix-tuning on Whisper for multitasking ASR and zero-shot E2E SLU. A novel semantic question generation method is also described.

\subsection{Whisper model}

Whisper is an open-sourced speech processing model developed by OpenAI, trained on 68000 hours of multilingual and multitask supervised data sourced from the web. This diverse dataset contributes to enhanced robustness, enabling Whisper to effectively handle variations in accents, background noise, and technical language \cite{radford2023robust}. Whisper follows a simple E2E architecture that builds upon an encoder-decoder Transformer. The encoder processes the log-Mel spectrogram of a 30-second audio chunk as input. Depending on the resulting speech embeddings and user-provided special tokens, the decoder executes tasks including speech transcription, language identification, and \textbf{X}-to-English translation. The decoding prompt may even incorporate transcription history and user instructions to influence the model output \cite{peng2023prompting}. With its versatility, the Whisper decoder is recognised as a robust audio-conditioned language model, making it a suitable choice for zero-shot SLU.

\subsection{Prompt-based QA framework for SLU}
\label{ssection:qa}

Given a spoken utterance $\mathbf{X}$, our goal is to: (1) identify its intent from a set of candidates $\mathbf{I}$ and (2) find all the entities in $\mathbf{X}$ corresponding to a set of slot types $\mathbf{S}$. In conventional supervised learning methods, these tasks are usually treated as utterance or word-level classification problems \cite{chen2019bert}. However, this approach lacks flexibility in accommodating new elements introduced to $\mathbf{I}$ and $\mathbf{S}$ without further training the classifier, a scenario typically known as zero-shot learning. 

%For example, an intent question \emph{``Does the user want to play\ music''} for the intent label {\fontfamily{qcr}\selectfont{play\_music}}, and a slot question \emph{``What is the genre of music?''} for the intent label {\fontfamily{qcr}\selectfont{music\_genre}}. 
In this work, we propose ZS-Whisper-SLU, a generative question-answering (QA) framework based on Whisper, to conduct zero-shot intent classification and slot filling (see Figure \ref{fig:zs-whisper-slu}). Descriptive questions are created for each element in $\mathbf{I}$ and $\mathbf{S}$ to construct question sets $\mathbf{Q_I}$ and $\mathbf{Q_S}$. The tasks are executed in three stages as follows:
\begin{enumerate}
    \item ASR: We adhere to Whisper's standard transcription procedure and reserve the ASR decoder states in a cache. These states encompass both acoustic features from the encoder and textual dependencies captured at the decoder side.
    \item Intent classification: The 1-best ASR transcript from stage 1 is duplicated and combined with all questions in $\mathbf{Q_I}$. The resulting sequences (intent prompts) are organised into a mini-batch and directly input to the Whisper decoder to generate binary answers, \emph{``Yes''} or \emph{``No''}, in parallel. Both answers, in plain texts, are tokenised using Whisper's own vocabulary, so that no additional verbaliser (binary classifier) is needed. As the intent questions in these text prompts do not have paired speech embeddings, the cross-attention module at each decoder layer is skipped \cite{li2023towards}. Apart from the ASR transcript in the prompt, the answer generation also relies on the cached ASR states for the purpose of E2E modelling.
    \item Intent-restricted slot filling: The decoding process is similar to intent classification. However, instead of feeding all questions in $\mathbf{Q_S}$ to the Whisper decoder, we select those relevant to the intent identified in stage 2. This not only reduces computation costs but also eliminates distracting slot types (e.g. the slot type {\fontfamily{qcr}\selectfont{artist}} may not be pertinent to the intent {\fontfamily{qcr}\selectfont{takeaway\_order}}). The slot answers are open-ended and correspond to the entities mentioned in $\mathbf{X}$. If an entity does not exist for a certain slot type, the system is supposed to produce an \emph{[EOT]} (end of transcript) symbol alone.
\end{enumerate}

Given the potential size of $\mathbf{Q_I}$ and $\mathbf{Q_S}$, we employ contrastive learning on each utterance for efficient training. This involves creating a mini-batch by combining positive questions representing the ground-truth intent and slots, along with $N$ randomly-sampled negative questions for both semantics \cite{sun2023knowledge}. The system is optimised using cross-entropy loss, with prepared answers serving as labels. During inference, the entire $\mathbf{Q_I}$ set is prompted to Whisper, and the question with the highest likelihood of \emph{``Yes''} answer is picked as the predicted intent. Non-empty answers to the chosen slot questions are collected as entity outputs. In cases where multiple slot types yield the same answer, the one with the highest probability is selected. Within such a framework, one can easily introduce new intents and slot types by adding their questions to $\mathbf{Q_I}$ and $\mathbf{Q_S}$, respectively.

\subsection{Prefix-tuning}
\label{ssection:prefix-tuning}

Prefix-tuning \cite{li2021prefix, liu2021p} is a widely used parameter-efficient fine-tuning (PEFT) technique for adapting deep neural networks such as LLMs and computer vision (CV) models. Recent works have expanded its application to speech-related tasks and yielded promising outcomes \cite{chang2023speechprompt}. Despite optimising a minimal amount of additional parameters (typically less than 1\%) to the existing model, prefix-tuning could achieve superior adaptation effects compared to the traditional full fine-tuning strategies.

As illustrated in Figure \ref{fig:zs-whisper-slu}, to steer Whisper towards producing reliable ASR transcripts and desired SLU answers, we concatenate a sequence of trainable prefix vectors with the speech/text embeddings at each encoder/decoder layer. Specifically, these vectors are injected to the key $\mathbf{K}$ and value $\mathbf{V}$ elements involved in Transformer's self-attention module:
\begin{equation}
    \mathbf{K} := \mathrm{Concat}(\mathbf{p^K}, \mathbf{K}),
    \mathbf{V} := \mathrm{Concat}(\mathbf{p^V}, \mathbf{V}),
\end{equation}
where $\mathbf{p^K}$ and $\mathbf{p^V}$ denote the prefix vectors generated by a prefix encoder (implemented as an embedding layer), serving as the soft contextual information for the ongoing encoding/decoding process. Note that the prefix vectors are shared across ASR and SLU tasks at each decoder layer, facilitating the exchange of shared knowledge such as entity names.

In our proposed system, the prefix vectors play three key roles: (1) adapting Whisper to the acoustic characteristics of the SLU corpus, (2) enhancing language modelling for task-specific domains, and addressing text normalisation issues like capitalisation and punctuation in ASR transcripts, and (3) guiding the Whisper decoder to produce intent and slot answers conditioned on the ASR decoder states. Moreover, the employment of prefix-tuning fully reserves the capabilities of the original Whisper model, ensuring flexibility for other use cases.

\subsection{SLU question generation}
\label{ssection:question}
High-quality questions are essential for accurately extracting semantic arguments within QA-based NLU/SLU frameworks. Common question design methods in prior research involve: i) adopting handcrafted templates \cite{lamanov2022template} and ii) deriving questions from raw label names \cite{li2023generative, sun2023knowledge}. However, these methods are prone to human biases and suffer from poor generalisation. 
For instance, the SLURP corpus includes an intent class labeled {\fontfamily{qcr}\selectfont{convert\_datetime}}, where users seek to convert date or time between different time zones. It is obvious that the true intent is not explicitly conveyed in the label name. Transforming such names directly into semantic questions may produce misleading prompts, adversely impacting SLU performance.

\begin{figure}[t]
\centering
  \centerline{\includegraphics[width=0.45\textwidth]{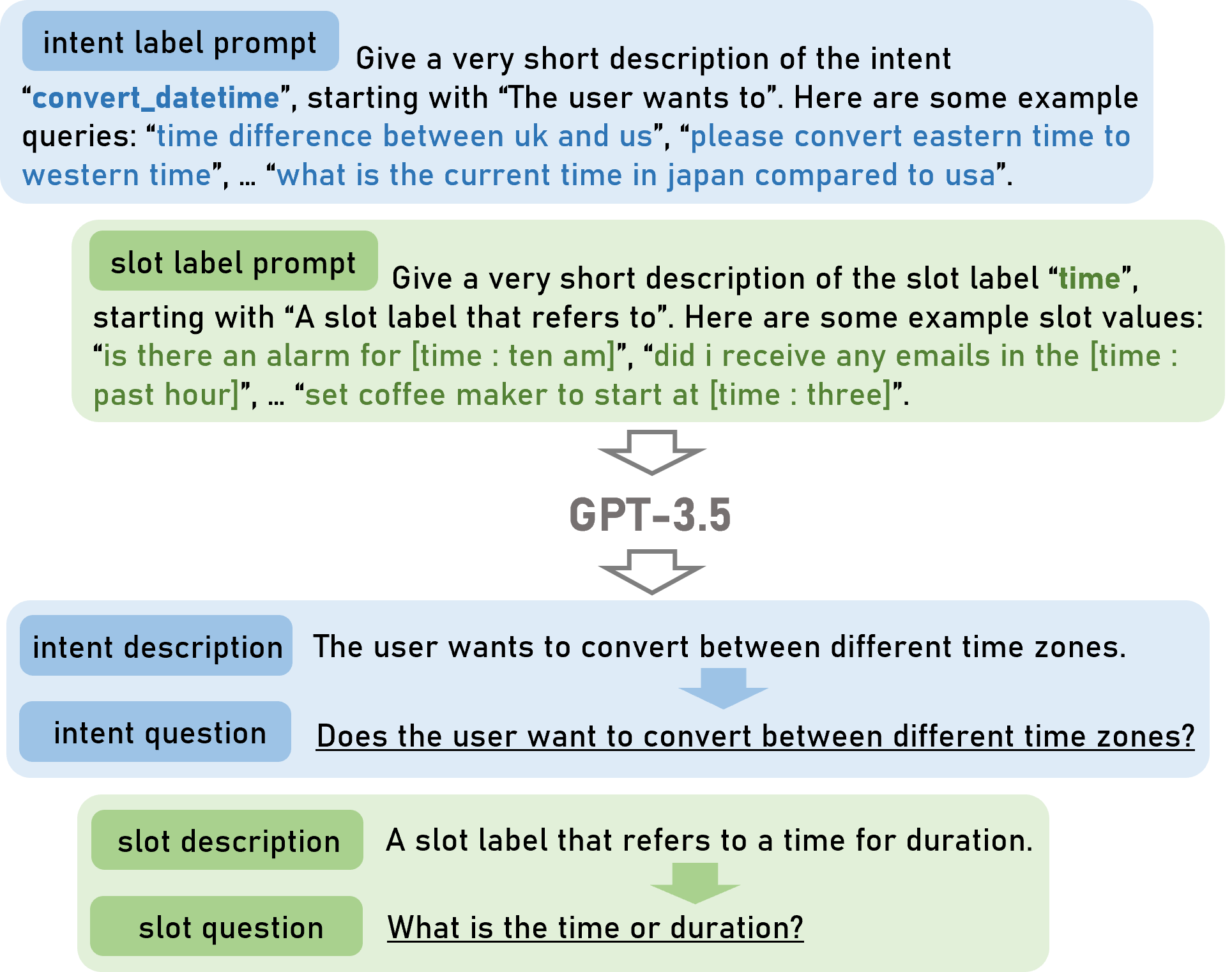}}
\caption{An example of LLM-based question generation process for intent and slot labels.}
\label{fig:question}
\vspace{-3mm}
\end{figure}

Consequently, we leverage LLMs such as GPT-3.5 \cite{brown2020language} to generate semantic questions, providing both the label name and a few user utterance examples to mitigate naming biases. We first instruct the LLM to produce a concise description for each semantic label, using the prompt templates presented in Figure \ref{fig:question}. Then the output description is translated into a question of a fixed format. We observe enhanced stability in adopting such a two-step question generation process, compared to prompting the LLM for a single-shot question production.

\begin{comment}
\vspace{1mm}

\emph{Give a very short description of the intent ``[intent\_label]'', starting with "The user wants to". Here are some example queries: ``[example\_1]'', ``[example\_2]'', ..., ``[example\_N].''}

\vspace{1mm}

\noindent for intent labels, and 

\vspace{1mm}

\emph{Give a very short description of the slot label ``[slot\_label]'', starting with ``A slot label that refers to''. Here are some example slot values: ``[example\_1]'', ``[example\_2]'', ..., ``[example\_N].''} 

\vspace{1mm}

\noindent for slot labels. Then the resulted description is translated into the semantic-informed question, in the format of:

\vspace{1mm}

\emph{Does the user want to [intent\_description]?}

\vspace{1mm}

\noindent for intent labels, and 

\vspace{1mm}

\emph{What is the [slot\_description]?} 

\vspace{1mm}

\noindent for slot labels. 
\end{comment}

\section{Experiments}
\label{section:exp}

\subsection{Datasets}

We evaluate the proposed system on three datasets:

\vspace{1mm}

\noindent \textbf{SLURP} \cite{bastianelli2020slurp}: The official SLURP dataset includes 65 intents and 56 slots within the domain of in-home personal robot assistant. Following \cite{sun2023knowledge}, a new data split is created for a zero-shot evaluation on 5 randomly selected slot types {\fontfamily{qcr}\selectfont\{podcast\_name, artist\_name, audiobook\_name, business\_name, radio\_name\}}. This adapted dataset is utilised for our in-corpus evaluation.

\vspace{1mm}

\noindent  \textbf{FSC} \cite{lugosch2019speech}: A widely used voice assistant corpus, featuring spoken commands labelled with {\fontfamily{qcr}\selectfont\{action, object, location\}} arguments. The argument values are drawn from a predefined set and constitute a total of 31 intents. We reorganise them into 15 intents and 2 slots ({\fontfamily{qcr}\selectfont\{language, location\}}, in order to accommodate the [\emph{None}] value), and transform the results back to the original format for evaluation. This dataset is used for our cross-corpus experiments.

\vspace{1mm}

\noindent  \textbf{Smartlight} (close-field) \cite{saade2019spoken}: A subset of the SNIPS SLU benchmark that contains instructions for controlling smart lights. The dataset comes with 6 intents and 3 slots to cover the functions and attributes of lights. In addition to FSC, we incorporate SmartLight into the cross-corpus evaluations.

\begin{table*}[hbt!]
\centering
\caption{\textbf{In-corpus evaluation.} Word error rate (WER \%) and slot filling SLU-F1 (\%) on the zero-shot test set of SLURP. The system architecture, trainable and total number of parameters of the systems are also provided. $\dagger$ indicates components in the E2E system.}
\vspace{-2mm}
\begin{tabular}{lccccc}
\hline \thickhline
    \multicolumn{1}{l}{Model} & Architecture & \#Trainable Param & \#Total Param & WER & SLU-F1 \\[1pt] \thickhline \midrule
    \multirow{2}{*}{Modular \cite{sun2023knowledge}} &\multicolumn{1}{c}{ASR: Conformer-LSTM} &\multirow{2}{*}{$\approx$220M} &\multirow{2}{*}{$\approx$220M} &\multirow{2}{*}{-} &\multirow{2}{*}{10.1} \\
        &\multicolumn{1}{c}{NLU: GPT-2-small} & & & & \\ \midrule
    \multirow{2}{*}{KA2G \cite{sun2023knowledge}} &\multicolumn{1}{c}{ASR$^\dagger$: Conformer-LSTM} &\multirow{2}{*}{$\approx$220M} &\multirow{2}{*}{$\approx$220M} &\multirow{2}{*}{18.0} &\multirow{2}{*}{9.3} \\
        &\multicolumn{1}{c}{NLU$^\dagger$: GPT-2-small} & & & & \\ \midrule
    \multirow{2}{*}{Modular} &\multicolumn{1}{c}{ASR: Whisper-large-v2} &\multirow{2}{*}{3.3M} &\multirow{2}{*}{2.3B} &\multirow{2}{*}{8.5} &\multirow{2}{*}{43.5}  \\
        &\multicolumn{1}{c}{NLU: GPT-2-large} & & & & \\ \midrule
    \multicolumn{1}{l}{\textbf{ZS-Whisper-SLU (proposed)}} &\multirow{3}{*}{Whisper-large-v2} &\multirow{3}{*}{3.3M} &\multirow{3}{*}{1.5B} &\multicolumn{1}{c}{\textbf{8.3}} &\multicolumn{1}{c}{\textbf{50.0}} \\
        \multicolumn{1}{l}{\quad\quad - ASR transcript} & & & &\multicolumn{1}{c}{8.4} &\multicolumn{1}{c}{45.8} \\
        \multicolumn{1}{l}{\quad\quad - ASR states} & & & &\multicolumn{1}{c}{8.5} &\multicolumn{1}{c}{36.3} \\[2pt] \hline \thickhline
\label{tab:slurp}
\end{tabular}
\vspace{-4mm}
\end{table*}

\begin{table}[t]
\centering
\caption{\textbf{Cross-corpus evaluation.} WER (\%) and intent classification accuracy (Acc. \%) on the test set of FSC. Systems with $*$ are supervised trained with labelled data.}
\vspace{-2mm}
%\resizebox{\columnwidth}{!}{%
\begin{tabular}{lcc}
\hline \thickhline
\multicolumn{1}{l}{Model} &\multicolumn{1}{c}{WER} &\multicolumn{1}{c}{Acc.} \\ \thickhline \midrule
\multicolumn{1}{l}{Whisper-SLU* \cite{meeus2023whisper}} &\multicolumn{1}{c}{-}  & \multicolumn{1}{c}{99.7} \\ \midrule
\multicolumn{1}{l}{Modular} &\multicolumn{1}{c}{0.8} &\multicolumn{1}{c}{95.1} \\ \midrule
\multicolumn{1}{l}{ZS-Whisper-SLU} &\multicolumn{1}{c}{0.8} &\multicolumn{1}{c}{95.0}  \\[2pt]
\hline \thickhline
\label{tab:fsc}
\end{tabular}%
%}
\vspace{-5mm}
\end{table}

\begin{table}[t]
\centering
\caption{\textbf{Cross-corpus evaluation.} WER (\%), intent classification accuracy (Acc. \%), slot filling SLU-F1 (\%) and perfect-parsing (PP \%) on the full set of SmartLight. Systems with $*$ are supervised trained with labelled data.}
\vspace{-2mm}
%\resizebox{\columnwidth}{!}{%
\begin{tabular}{lcccc}
\hline \thickhline
\multicolumn{1}{l}{Model} &\multicolumn{1}{c}{WER} &\multicolumn{1}{c}{Acc.} &\multicolumn{1}{c}{SLU-F1} &\multicolumn{1}{c}{PP} \\ \thickhline \midrule
\multicolumn{1}{l}{Finstreder* \cite{bermuth2022finstreder}} &\multicolumn{1}{c}{6.1} &\multicolumn{1}{c}{-} &\multicolumn{1}{c}{-} &\multicolumn{1}{c}{88.0} \\ \midrule
\multicolumn{1}{l}{Whisper-SLU* \cite{meeus2023whisper}} &\multicolumn{1}{c}{-}  &\multicolumn{1}{c}{95.4} &\multicolumn{1}{c}{-} &\multicolumn{1}{c}{-} \\ \midrule
\multicolumn{1}{l}{Modular} &\multicolumn{1}{c}{2.8} &\multicolumn{1}{c}{91.9} &\multicolumn{1}{c}{90.8} &\multicolumn{1}{c}{82.0} \\ \midrule
\multicolumn{1}{l}{ZS-Whisper-SLU} &\multicolumn{1}{c}{2.7} &\multicolumn{1}{c}{91.6} &\multicolumn{1}{c}{90.9} &\multicolumn{1}{c}{82.5} \\[2pt]
\hline \thickhline
\label{tab:smartlight}
\end{tabular}%
%}
\vspace{-6mm}
\end{table}

\subsection{Experimental setups}

We choose {\fontfamily{qcr}\selectfont Whisper-large-v2} (with 1.5 billion parameters) as the foundation ASR model for zero-shot SLU adaptation. The proposed system is compared against three baselines:
\begin{itemize}
    \item A Modular system in \cite{sun2023knowledge}: Comprising a Conformer-LSTM-based ASR module and a GPT-2-based \cite{radford2019language} NLU module.
    \item KA2G \cite{sun2023knowledge}: The E2E counterpart to the above modular structure, with the ASR and NLU components connected through a neural aligner at the decoder state level.
    \item A Modular system developed in-house: Consisting of a Whisper ASR module and a GPT-2 NLU module. For a fair comparison, the ASR module also utilises {\fontfamily{qcr}\selectfont Whisper-large-v2}, and the NLU module employs {\fontfamily{qcr}\selectfont GPT-2-large} (with 0.8 billion parameters), which shares a similar architecture with the ASR decoder. Both modules are prefixed-tuned using the same hyper-parameters as for the proposed system.
\end{itemize}
We do not compare our ZS-Whisper-SLU system with the audio-to-intent approach in \cite{elluru2023generalized} due to undisclosed data splits and enrollment signals not utilised by our system.

For the in-corpus experiments on SLURP, both ZS-Whisper-SLU and the GPT-2 module in the modular system are trained using $N=10$ negative intent and slot examples for each utterance. The prefix length is set to 10 per task, resulting in 10 prefix vectors at each encoder layer (for ASR) and 30 prefix vectors at each decoder layer (for joint ASR, intent classification, and slot filling). All models undergo 10 epochs of training using the AdamW optimiser with a mini-batch size of 12. A linear learning rate scheduler, starting from 0.002 without warmup, is employed for weight decay. The SLURP-developed models are directly applied to the cross-corpus evaluations on FSC and SmartLight. However, the encoder prefix vectors are disabled to prevent transfers of acoustic knowledge. We utilise the official test set of FSC and all utterances in SmartLight, since it lacks a data split. For fast inference, all transcripts and semantic answers are generated with greedy-search decoding.

\subsection{Experimental results}

Table \ref{tab:slurp} presents the in-corpus ASR and slot filling results on the zero-shot test set of SLURP. It is observed that the proposed ZS-Whisper-SLU system outperforms both the modular and E2E KA2G systems from \cite{sun2023knowledge} in both metrics. Thanks to the ASR power of Whisper, our system achieves a substantial 9.7\% reduction in word-error-rate (WER) compared to KA2G. This further leads to an absolute 40.7\% improvement in SLU-F1 score \cite{bastianelli2020slurp} on the unseen slots. ZS-Whisper-SLU also demonstrates superiority over a much stronger baseline, our own modular system that incorporates a GPT-2-large LLM for NLU, with a relative SLU-F1 gain of 14.9\%. In addition to enhanced performance, ZS-Whisper-SLU utilises significantly fewer parameters (34.8\% relative reduction) than the Whisper-GPT-2 modular system, highlighting the robust language understanding capabilities of the Whisper decoder and the efficacy of E2E modelling. From a training efficiency perspective, the zero-shot SLU functionalities in the proposed system are achieved by optimising a prefix encoder with only 3.3 million parameters, which constitutes just 0.2\% of the original Whisper model.

%ASR performance is measured by word error rate (WER \%), while slot filling is assessed using SLU-F1 (SF SLU-F1 \%) score \cite{bastianelli2020slurp} to consider the influence of ASR errors. The trainable and total number of parameters for each system are also included.

The cross-corpus zero-shot evaluation results for FSC and SmartLight are displayed in Table \ref{tab:fsc} and \ref{tab:smartlight}, respectively. To illustrate the robustness of ZS-Whisper-SLU, we also compared it with the state-of-the-art supervised training systems from the literature on both datasets. The results indicate that the proposed system consistently achieves competitive performance with the modular structure. Notably, it performs comparably well to the supervised models, showing a slight difference of 4.7\% on FSC and 5.5\% (PP) on SmartLight, despite not being exposed to any in-corpus data during training. This underscores the ability of ZS-Whisper-SLU to effectively transfer knowledge from seen domains to unseen domains.

\subsection{Ablation study}

The ablation study results on SLURP are presented in the final row of Table \ref{tab:slurp}. To assess the key factor influencing ZS-Whisper-SLU's zero-shot performance, we selectively prune certain components during the QA decoding process:
\begin{itemize}
    \item Removal of ASR transcript in the question prompt: In this case, decoding relies solely on the intent/slot question and cached ASR states generated during transcription.
    \item Exclusion of ASR states: This aligns with the modular approach, where the Whisper decoder functions as an independent NLU module, processing text-based ASR transcripts and semantic questions in the prompts. 
\end{itemize}
A significant performance drop of 13.7\% in SLU-F1 is observed when excluding ASR states from the decoding history in the proposed system. This operation undermines the E2E decoding property of the Whisper model, thus leading to inferior behaviour. Omitting ASR transcript from the prompt results in a moderate SLU-F1 decrease (4.2\%), which suggests that including the text stabilises semantics extraction by mitigating acoustic variance embedded in the ASR states. Therefore, employing joint modelling on speech and text modalities achieves optimal zero-shot performance for ZS-Whisper-SLU.

\section{Conclusions}
\label{section:con}
This paper presents the use of OpenAI's Whisper model for zero-shot E2E SLU. We address intent classification and slot filling tasks within a QA framework to handle unseen labels. The intent and slot values are derived by prompting questions to the Whisper decoder through prefix-tuning. Our experiments show that ZS-Whisper-SLU achieves competitive performance in both in-corpus and cross-corpus settings compared to a modular system, while using notably fewer model parameters.

\bibliographystyle{IEEEtran}
\bibliography{mybib}

\end{document}